\documentclass[aps,prc,twocolumn,nofootinbib,superscriptaddress,groupedaddress]{revtex4}
\usepackage{dcolumn}   
\usepackage{bm}        
\usepackage{multirow}
\usepackage{xcolor}

\usepackage{amsmath}
\usepackage{amsfonts}
\usepackage{amssymb}
\usepackage{makeidx}
\usepackage{graphicx}
\usepackage{anysize}
\usepackage{hyperref}
\usepackage{slashed}

\newenvironment{mymathbox}
{\par\smallskip\centering\begin{lrbox}{0}%
\begin{minipage}[c]{0.8\textwidth}}
{\end{minipage}\end{lrbox}%
\framebox[0.9\textwidth]{\usebox{0}}%
\par\medskip
\ignorespacesafterend}
\newcommand{\bb}{\begin{mymathbox}}
\newcommand{\eb}{\end{mymathbox}}
                                
\newcommand{\half}{\frac{1}{2}}
\newcommand{\mhalf}{\frac{-1}{2}}

\newcommand{\munit}{\mbox{\boldmath $1\!\!1$}}

\newcommand{\fl}[1]{\begin{flalign}#1\end{flalign}}

\newcommand{\be}{\begin{equation}}
\newcommand{\ee}{\end{equation}}
\newcommand{\ba}{\begin{eqnarray}}
\newcommand{\ea}{\end{eqnarray}}

\newcommand{\uuN}{u({\bf p}_N,s_N)}

\newcommand{\nk}{{\bf      k}}
\newcommand{\np}{{\bf      p}}
\newcommand{\nq}{{\bf      q}}
\newcommand{\nr}{{\bf      r}}

\newcommand{\npsi}{{\bf \npsi}}

\newcommand{\de}{\text{d}}

\newcommand{\non}{\nonumber}

\newcommand{\bma}{\begin{pmatrix}}
\newcommand{\ema}{\end{pmatrix}}

\begin{document}

\title{Nuclear effects in electron- and neutrino-nucleus scattering within a relativistic quantum mechanical framework}

\author{R.~Gonz\'alez-Jim\'enez}
\affiliation{Grupo de F\'isica Nuclear, Departamento de Estructura de la Materia, F\'isica T\'ermica y Electr\'onica, Facultad de Ciencias F\'isicas, Universidad Complutense de Madrid and IPARCOS, CEI Moncloa, Madrid 28040, Spain}
\author{A.~Nikolakopoulos}
\affiliation{Department of Physics and Astronomy, Ghent University, Proeftuinstraat 86, B-9000 Gent, Belgium}
\author{N.~Jachowicz}
\affiliation{Department of Physics and Astronomy, Ghent University, Proeftuinstraat 86, B-9000 Gent, Belgium}
\author{J.M.~Ud\'ias}
\affiliation{Grupo de F\'isica Nuclear, Departamento de Estructura de la Materia, F\'isica T\'ermica y Electr\'onica, Facultad de Ciencias F\'isicas, Universidad Complutense de Madrid and IPARCOS, CEI Moncloa, Madrid 28040, Spain}

\date{\today}

\begin{abstract}
We study the impact of the description of the knockout nucleon wave function on electron- and neutrino-induced quasielastic and single-pion production cross sections. We work in a fully relativistic and quantum mechanical framework, where the relativistic mean-field model is used to describe the target nucleus. The focus is on Pauli blocking and the distortion of the final nucleon, these two nuclear effects are separated and analyzed in detail. 
We find that a proper quantum mechanical treatment of these effects is crucial to provide the correct magnitude and shape of the inclusive cross section. Also, this seems to be key to predict the right ratio of muon- to electron-neutrino cross sections at very forward scattering
angles.
\end{abstract}


\maketitle


\section{Introduction}

The main goals of the new generation of accelerator-based neutrino oscillation experiments, DUNE~\cite{DUNE16} and T2HK~\cite{HyperK15}, are to measure the CP violating phase in the lepton sector, improve the accuracy on the value of the mixing angle $\theta_{23}$ and determine the neutrino mass ordering~\cite{Alvarez-Ruso18}.
One of the main problems in achieving the unprecedented level of accuracy required by these programmes is that the neutrino beams are not monoenergetic.
The reconstructed neutrino energy, the main ingredient for the oscillation analysis, is known as a broad distribution that ranges from tens of MeVs to a few GeVs.
This energy is reconstructed using Monte Carlo (MC) simulations that rely not only on the available experimental information in the detectors, but also crucially on the models for neutrino-nucleus interactions that are implemented in the MC event generators.

In recent years, important progress has been made on the experimental side to increase the statistics and to reduce systematic uncertainties, for instance, those related with the normalization of the flux~\cite{MINERvA-flux16}.
On the theoretical side, many studies have been presented aiming at improving our knowledge on lepton-nucleus scattering, and more specifically on neutrino-nucleus scattering~\cite{Katori17, Butkevich07, Nieves11, Ivanov13, Ivanov15, Martini16, Ankowski15, Gallmeister16, Rocco16, Lovato16, VanCuyck17, VanDessel17}. However, it is extremely challenging to provide reliable and consistent predictions of the diversity of processes that can take place in the energy range covered by the neutrino beams.
All this, together with the fact that the theoretical improvements, for one reason or another, do not always readily find their way to generators, has caused the neutrino-nucleus cross sections to become one of the major sources of uncertainties in the reconstruction of the neutrino energy and, as a consequence, in the oscillation analysis~\cite{Alvarez-Ruso18}.

Electron scattering studies over the last 50-60 years have allowed us to identify the main reaction mechanisms driving the lepton-nucleus interaction in the intermediate energy region~\cite{Foundations17}.
The nuclear response obtained in inclusive $(e,e')$ experiments is usually understood in terms of a few mechanisms: discrete and collective nuclear excitations, quasielastic (QE) scattering, multinucleon excitations, single-pion production (SPP), and deep-inelastic scattering (DIS). Though this is probably an oversimplified picture, it has proven to work reasonably well, as shown e.g. in Refs.~\cite{Megias16a, Gallmeister16, Ivanov16a, Rocco16}. 

In this work, we analyze nuclear effects that are important to understand two of the main reaction mechanisms in the intermediate energy region, namely, QE scattering and SPP.
We will show that the residual nucleus plays an important role in the scattering process via Pauli blocking and distortion of the outgoing nucleon. As will be shown, a proper quantum mechanical treatment of these two effects is crucial to provide the right magnitude and shape of the cross sections. As recently shown in Ref.~\cite{Nikolakopoulos18b}, this could influence the reconstruction of the neutrino energy.

\begin{figure}[htbp]
  \centering
      \includegraphics[width=.35\textwidth,angle=0]{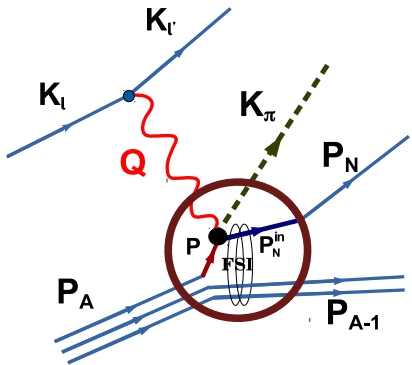}
  \caption{ Representation of the SPP process within our approach. Each particle is labeled by its 4-vector. The incoming lepton $K_l(\varepsilon_i,\nk_i)$ goes to $K_{l'}(\varepsilon_f,\nk_f)$ by exchange of a boson $Q(\omega,\nq)$. The boson couples to a bound nucleon $P(E,\np)$ in the target nucleus $P_A(E_A,\np_A)$. The nuclear volume where the interaction occurs is represented by the red circle. Inside the nuclear volume the struck nucleon $P_N^{in}$ is affected by the residual nucleus $P_{A-1}(E_{A-1},\np_{A-1})$. Outside the nuclear volume one finds the nucleon $P_N(E_N,\np_N)$ and the pion $K_\pi(E_\pi,\nk_\pi)$. 
  The same picture is valid for the QE reaction but without the pion line.
}
  \label{fig:diagram}  
\end{figure}

Our models are based on the impulse approximation, that is, the lepton interacts only with the knockout nucleon in the nucleus; and the first-order Born approximation, i.e., only one boson is exchanged between the lepton and the hadron system. This is depicted in Fig.~\ref{fig:diagram}. 
What makes our approach attractive and different from others commonly used in neutrino nucleus scattering is that we work in a fully relativistic and quantum mechanical framework:
\begin{itemize}
 \item Relativity: 
The degrees of freedom in a relativistic and a non-relativistic calculations are not fully equivalent.
Some ingredients of a relativistic calculations mimick the effect of many-body interactions in a non-relativistic calculation~\cite{Serot97}. For instance, in the RMF model, contrary to non-relativistic ones, there is no need of NN or NNN correlations to get the saturation minimum for nuclear matter.
Related to this, the spectroscopic factors extracted within independent-particle shell models from  exclusive $(e,e'p)$ experiments are systematically larger in relativistic models than those in non-relativistic ones ($\sim0.75$ compared to $\sim0.55$ for $^{40}$Ca and $^{208}Pb$)~\cite{Udias95}. 
While nuclei are certainly strongly correlated systems, phenomenological mean field models could effectively incorporate some of these correlations already at the mean field level. The exclusive nature of these experiments will maximize the overlap of the measurement with the mean field prediction. The depletion of the spectroscopic factor indicates the importance of the residual part of the interaction, not taken into account at the mean field level. The larger spectroscopic factors derived from the relativistic models seem to suggest that these models have a better capacity to effectively incorporate correlations within the mean field calculation.
Finally, it is worth noting that the relativistic-mean field model employed here~\cite{Walecka74,Horowitz81} has only 6 free parameters: coupling constants of the $\omega$, $\rho$ and $\sigma$ mesons and the mass of the later, and two additional coupling constants for non-linear terms.
 More complete discussions about relativistic effects and its relation to non-relativistic calculations can be found in~\cite{Serot97,Udias95,Amaro07} and references therein.

 In our framework, both kinematic and dynamic relativistic effects are naturally incorporated.  
  
 \item Quantum Mechanics: In a scattering process from a quantum mechanical system, such as the nucleus, the only available information is the initial and final states, nothing is known about the intermediate steps. 
 Therefore, the classical description of the process as a factorization of consecutive scatterings, e.g. the treatment of final-state interactions in a cascade model, is under some circumstances not constrained by reality. This is especially relevant when the wavelength of the scattering particle is larger than or of the same order of magnitude as the size of the target, for instance, for outgoing nucleons with low momenta.
As will be shown, when a low momentum nucleon is knocked out in the QE process one finds a very particular shape for the cross section which can be partially understood from the orthogonality of initial and final state wave functions, and is not reproduced in a factorized model.  
 
We perform a quantum mechanical calculation of the final-state interactions (FSI) by solving the Dirac equation for the outgoing nucleon, which is under the influence of relativistic mean-field potentials. 
\end{itemize}

It is important to point out that, to our knowledge, the distortion of the outgoing nucleon within the RMF model is implemented for the first time in a pion-production calculation. 
In Refs.~\cite{Fernandez-Ramirez08,Praet09}, and more recently in~\cite{Gonzalez-Jimenez18a,Nikolakopoulos18a}, the relativistic plane wave impulse approximation (RPWIA) was employed to make predictions on SPP on the nucleus. Within RPWIA, the initial state is described by the relativistic mean-field model while the pion and knocked out nucleon are treated as plane waves, i.e., FSI are totally neglected. 
This is obviously an extreme simplification of the problem. To amend that, as mentioned, we have included the distortion of the scattered nucleon in our SPP model. The effect of this will be shown along this work.
Finally, we want to point out that we do not include the distortion or absorption of the pion, work in that direction is in progress.

In Sect.~\ref{Models}, we discuss the different approaches employed in this work to describe the outgoing nucleon wave function.
Results for electron and neutrino QE and SPP cross sections are shown and analyzed in Sect.~\ref{Results}. Our conclusions are presented in Sect.~\ref{Conclusions}.

\section{Models}\label{Models}

We describe QE scattering cross section as:
\fl{
 \frac{\de^{5}\sigma}{\de \varepsilon_f\de\Omega_f \de\Omega_N} = {\cal F} \frac{p_NE_N k_f \varepsilon_f}{(2\pi)^5f_{rec}} l_{\mu\nu}h_{QE}^{\mu\nu}\,,\label{QEXS}
}
with $f_{rec}=\left|1 + \frac{E_N}{p_N^2E_{A-1}}\np_N\cdot(\np_N-\nq) \right|$. The factor ${\cal F}$ and the leptonic tensor $l_{\mu\nu}$, which depend on the type of interaction (electromagnetic, charged current, or weak neutral current), were defined in~\cite{Gonzalez-Jimenez17}. The kinematical variables are introduced in Fig.~\ref{fig:diagram}.

For the SPP process, represented in Fig.~\ref{fig:diagram}, we work with the cross section~\cite{Gonzalez-Jimenez18a,Gonzalez-Jimenez18b}:
\footnote{In Eqs.~\ref{QEXS} and \ref{SPPXS}, the degree of freedom linked to the excitation energy of the residual system has already been integrated out. 
In our shell model, the missing energy $E_m$, defined as the part of the energy transferred $\omega$ that transforms into internal energy of the residual system, is a constant value for each shell. This produces an energy conservation Dirac delta that can be trivially integrated. We have checked that the replacement of this delta function by a distribution does not introduce any significant effect in the inclusive cross sections studied here.} 
\fl{
 \frac{\de^{8}\sigma}{\de\varepsilon_f\de\Omega_f \de E_\pi \de\Omega_\pi \de\Omega_N} =& {\cal F} \frac{k_f \varepsilon_f p_N E_N E_\pi k_\pi}{(2\pi)^{8}f_{rec}}\nonumber\\
 \times& l_{\mu\nu}h_{SPP}^{\mu\nu}\label{SPPXS}
}
with $f_{rec} = \left|1 + \frac{E_N}{E_{A-1}}\left(1 +\frac{\np_N\cdot(\nk_\pi-\nq)}{p_N^2}\right)\right|$.

More inclusive results, e.g. $(e,e')$ cross sections, are obtained by summing in Eqs.\ref{QEXS} and \ref{SPPXS} over all occupied shells and integrating over the variables of the undetected particles.

The hadronic tensor for the scattering off a nucleon from a given shell is given by
\fl{ h^{\mu\nu}_X=\frac{1}{2j+1}\sum_{m_j,s_N} (J_X^\mu)^\dagger J_X^\nu\,,
}
where $X$ denotes the type of process (QE or SPP), $j$ the total angular momentum of the bound nucleon, $m_j$ its third component, and $s_N$ the spin projection of the outgoing nucleon. 
We average over initial bound states for a given shell ($\frac{1}{2j+1}\sum_{m_j}$) and sum over final states ($\sum_{s_N}$).
 
\begin{figure}[htbp]
  \centering
      \includegraphics[width=.25\textwidth,angle=270]{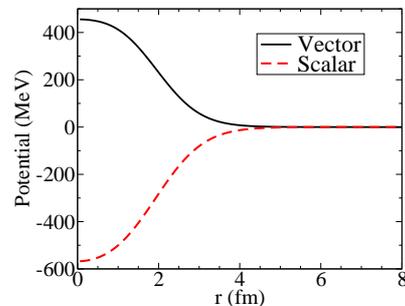}
      \vspace{-0.3cm}
      \caption{ RMF vector and scalar potentials as a function of the position $r$ in the $^{12}C$ nucleus. }
  \label{fig:potentials} 
\end{figure}

In coordinate space, the hadronic current $J_X^\mu$ reads
\ba
J^\mu_X = {\cal C}\int_V \de\nr\ \overline{\Psi}^{s_N}(\nr,\np_N)\ {\cal O}_X^\mu \ e^{i\nq\cdot\nr}  \psi_\kappa^{m_j}(\nr)\,,
\ea
with $V$ the nuclear volume and ${\cal C}$ the coupling constant of the hadronic vertex, defined in Ref.~\cite{Gonzalez-Jimenez17}.
All along this work, the bound state wave function $\psi_\kappa^{m_j}(\nr)$, labeled with the quantum numbers $\kappa$ and $m_j$~\cite{Caballero98a}, is always computed in the same way, i.e., within the RMF model~\cite{Serot97,Horowitz81}.
This accounts for Fermi motion and binding energy in a consistent way. 
$\Psi^{s_N}(\nr,\np_N)$ is the wave function of the outgoing nucleon which has asymptotic momentum $\np_N$ and spin projection $s_N$.
In Fig.~\ref{fig:potentials} we represent the vector and scalar RMF potentials used in our calculations. Notice that the Coulomb potential that affects the nucleons is included in our calculations but not in the figure.

The transition between the initial and the final state is given by the relativistic operator ${\cal O}_X^\mu$. For QE scattering we use the usual CC2 operator (see e.g. Ref.~\cite{Caballero06}). 
For SPP we use the operator described in~\cite{Gonzalez-Jimenez17}, that contains the delta, D13, S11 and P11 resonances, and background terms (first-order contributions of the $\chi$PT Lagrangian for the pion-nucleon system~\cite{Hernandez07}). At large invariant mass ($W$), the background terms are replaced by a Regge inspired operator that provides the correct $W$ behavior of the amplitude~\cite{Gonzalez-Jimenez17}.

The goal of this work is to illustrate how the interaction between the outgoing nucleon and the residual nucleus affects the predicted QE and SPP cross sections. For that, we present results from several approaches corresponding to different treatments of the final state nucleon $\Psi^{s_N}(\nr,\np_N)$.
These are described in what follows.

\subsection{RPWIA model}

The outgoing nucleon is a relativistic plane wave. A well-known problem of this model, when the initial nucleus is described by any realistic nuclear model beyond the free Fermi gas, is that in this approach the orthogonality between initial and final nuclear states is not fulfilled~\cite{Boffi82,Atti83}. This is, within RPWIA one has
\ba
  \langle I|\munit|F\rangle \neq \delta_{I,F}, \label{prob_cons}
\ea
with $|I,F\rangle$ the full (leptonic and hadronic) initial and final states, and $\munit$ the unit operator.
This is due to the fact that the initial and final hadronic states are eigenstates of different hamiltonians: the free one for the outgoing nucleon and the RMF for the bound state, in our case. Non-orthogonality effects would be more conspicuous when the momentum of the initial and final nucleon are close, and thus they are irrelevant for high values of the momentum of the final nucleon~\cite{Boffi82,Johansson00}.
Indeed, the pathologies related to non-orthogonality  will show up in the cross sections when the momentum of the outgoing nucleon $p_N$ is small, i.e., when the overlap between the initial and final state wave functions is important. This will be shown in Sect.~\ref{Results}.

\subsection{Pauli-blocked RPWIA model} 

With this model, the idea is to make the initial and final states orthogonal; thus, restoring the equality in Eq.~\ref{prob_cons}.
To make the relativistic plane wave orthogonal to the initial state we define the wave function $\Psi^{s_N}(\nr,\np_N)$ for the knocked out nucleon~\cite{Johansson00}:
\fl{ 
  |\Psi^{s_N}(\np_N)\rangle &= |\psi^{s_N}_{pw}(\np_N)\rangle - \sum_{\kappa,m_j} [C^{m_j,s_N}_{\kappa}(\np_N)]^\dagger\, |\psi_{\kappa}^{m_j}\rangle\non \\
  &\times \Theta(\omega-E_m^{\kappa}-T_{A-1}) \,,\label{psi_block}
}
with $\psi^{s_N}_{pw}(\nr,\np_N)$ a plane wave with momentum $\np_N$ and spin projection $s_N$. 
The projection coefficient is:
\ba
C^{m_j,s_N}_{\kappa}(\np_N)\equiv\langle\psi^{s_N}_{pw}(\np_N)|\psi_\kappa^{m_j}\rangle\,.
\ea
The sum in Eq.~\ref{psi_block} runs over all bound states $\psi_{\kappa}^{m_j}(\nr)$. The analytic expressions of the projection coefficients are given in Appendix~\ref{proj-coeff}. In Appendix~\ref{normalization} we show that the wave function in Eq.~\ref{psi_block} is normalized to one.

Proper orthogonalization in the mean field formalism employed here means that all the single-particle states, both for the bound nucleons as well as the continuum states, should be orthogonal. This is automatically obtained if these single particle states, discrete and continuum, were computed with the same potential, as in the RMF approach (Sect.~\ref{RMFmodel}). When using free states for the final state, orthogonalization must be restored with a constructive Gram-Schmidt procedure~\cite{Johansson00}. We refer to it as PB-RPWIA.

On top of that, we have introduced the step function $\Theta(\omega-E_m^{\kappa}-T_{A-1})$ in Eq.~\ref{psi_block}. 
Here, $E_m^{\kappa}$ is the missing energy for the $\kappa$-shell and $T_{A-1}$ is the kinetic energy of the residual nucleus, 
which for the SPP process is given by 
$T_{A-1} = \omega - T_N - E_\pi - E_m$  (in the QE case, one simply removes $E_\pi$), with $T_N$ the kinetic energy of the outgoing nucleon and $E_m$ the missing energy.
This is done to ensure that only those contributions allowed by the kinematics are subtracted to the plane wave. 
For example, let us consider that $\omega=30$ MeV, then the nucleons from the $1s_{1/2}$-shell cannot be knocked out (the binding energy is $\sim40$ MeV). This means that for that kinematic, the initial state for our scattering problem is just made of the nucleons in the $1p_{3/2}$-shell,
so we do not subtract the $s$-wave contribution from the plane wave in the final state.
The introduction of this {\em ad hoc} prescription is guided by the better agreement with the ``full'', properly orthogonalized, RMF results (Sect.~\ref{RMFmodel}).
Some results computed within this approach were recently shown in Ref.~\cite{Nikolakopoulos19}.

The effects on the cross sections derived from the lack of orthogonality in the hadronic current have been previously studied for the one-nucleon knockout reactions $(\gamma,p)$ and $(e,e'p)$ (see~\cite{Boffi82,Atti83,Johansson00} and references therein). In particular, Eq.~\ref{psi_block} (without the step function), was used in Ref.~\cite{Johansson00}.

\subsection{RMF model} \label{RMFmodel}

The wave function of the knockout-nucleon is given by the solution in the continuum of the Dirac equation in the presence of the same RMF potentials used to obtain the bound state. In this way, orthogonality is ensured while distortion of the outgoing nucleon is accounted for by propagating it with the self energy computed within the mean field approach, with the same potential as for the bound nucleons. In a process for which nucleon propagation can be reasonably described by this model, such as inclusive reactions at moderate values of transferred momentum, this is expected to be a good approximation. It has been compared to $(e,e')$ data and other models in Refs.~\cite{Meucci09,Maieron03,Gonzalez-Jimenez14b,Barbaro19} and with CC neutrino-nucleus scattering data in Refs.~\cite{Maieron03,Amaro11a,Meucci11a,Ivanov13}. 
Indeed, in Ref.~\cite{Amaro07}, it was shown that the longitudinal scaling function of the RMF model follows the scaling behavior extracted by the analysis of experimental data~\cite{Maieron02}, while other non-relativistic shell models do not show this.

On the other hand, if the experiment obtains additional information, that is, it is not a fully inclusive experiment, we would need to account for effects such as absorption of the nucleons during propagation, explicit multi-nucleon emissions, other channels involving production of particles, etc. In the case of fully exclusive reactions, where the experiment selects just the elastic propagation of the nucleon in the residual nucleus, a phenomenological optical potential can be employed instead of the RMF to describe the nucleon propagation. This is done for instance, in Refs.~\cite{Udias93,Udias01} by using an optical potential with an imaginary part and correcting for the spectroscopic factors.

\subsection{Energy-Dependent RMF potentials}
 
Even for inclusive experiments, the pure RMF approach where the nucleon in the final state is described with an energy-independent potential is bound to fail as the momentum of the final nucleon increases. Unitarity demands that the potentials get softer with increasing nucleon energies. Indeed, in a consistent treatment of bound and scattering states in the framework of dispersion-relation approaches~\cite{Mahaux86,Mahaux91,Dickhoff12} or, in general, non-local schemes, one can build a mean-field potential smoothly evolving from a purely real strong potential for the bound states, and, as the energy increases, weaker and weaker potentials, with increasing imaginary contributions, depending on the actual scattering states considered. These approaches may preserve orthogonality, unitarity and dispersion relations.
In order to describe inclusive processes, the flux lost in inelasticities can be summed-out in every channel by means of Green's function approaches~\cite{Capuzzi91,Meucci03,Meucci05,Ivanov16b}.

Thus, it comes to no surprise that when the energy-independent RMF is employed to describe FSI at every kinematics, when the momentum of the knockout nucleon reaches $400-500$ MeV, the QE cross sections computed within the RMF model depart from the $(e,e')$ data~\cite{Gonzalez-Jimenez14b,Barbaro19}. In particular, one observes too much reduction of the QE peak, with the strength moved to the high-$\omega$ tail, and a large shift of the distributions towards higher $\omega$ values.

Based on this idea, the SuSAv2 model, presented in Ref.~\cite{Gonzalez-Jimenez14b}, builds a trade-off between RMF and RPWIA results. It uses a linear combination of the RPWIA and RMF scaling functions, with the weight of each contribution given by a transition function that depends on the momentum transfer $q$. In Ref.~\cite{Megias16a}, the parameters of the SuSAv2 were fitted to $^{12}C(e,e')$ data. The agreement with inclusive data achieved by the SuSAv2, for $\omega$ values above approximately 50 MeV is outstanding, including the comparison with $(e,e')$ data from different nuclei~\cite{Megias18,Barbaro19} and neutrino-induced reactions~\cite{Megias16b}.

\begin{figure}[htbp]
  \centering
      \includegraphics[width=.25\textwidth,angle=270]{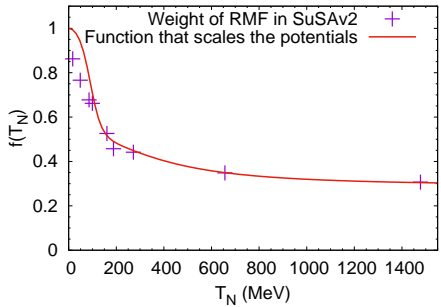}
      \caption{ Function that scales the RMF potentials. The crosses are addapted from Ref.~\cite{Megias16a}.}
  \label{fig:function} 
\end{figure}

Inspired by the success of this approach, here we introduce an energy-dependent potential that keeps the RMF strength and proper orthogonalization for slow nucleons, but the potential becomes softer for increasing nucleon momenta. Instead of microscopic calculations to derive the evolution of the potential with energy, we rely on the comparison to inclusive electron scattering data, as done in the SuSAv2 fit.
For that, we scale the scalar and vector RMF potentials of Fig.~\ref{fig:potentials} by multiplying them with the function $f(T_N)$ shown in Fig.~\ref{fig:function}. 
This function results from a fit to the weight of the RMF contribution in the SuSAv2 model (see Fig.~10 in Ref.~\cite{Megias16a}).~\footnote{In Fig.~10 of Ref.~\cite{Megias16a} the relative weight of RMF and RPWIA contributions in the SuSAv2 are shown as a function of $q_{QE}$, i.e. $q$ at the QE peak. To make the transformation from $q_{QE}$ to $T_N$ we have used the relation:
\ba
  \omega_{QE} \approx \sqrt{q_{QE}^2 + M^2} - M\,.\non
\ea
Then, we get $T_N \approx \omega -20$ MeV, where we have neglected the nuclear recoil and approximated $\omega$ by $\omega_{QE}$. The $-20$ MeV is to account for the average binding energy.
}
Notice that $f(T_N)$ follows SuSAv2 behavior (purple crosses) from approximately $T_N\gtrsim100$ MeV. For $T_N<100$ MeV, the RMF agrees with data well, therefore, we make the function $f(T_N)$ close to one in this region. In this way, we also avoid orthogonality issues that are especially relevant in this low momentum region. 

We stress that these new energy dependent potentials only influence the scattered nucleon, the initial state is described by the original RMF model. For simplicity, we chose to introduce the same energy dependence for the RMF potentials for protons and neutrons. The actual potentials, computed from the RMF would of course be different for protons and neutrons, and even more so for non isoscalar targets, but the energy dependence of both potentials can safely assumed to be very similar. This is supported by other phenomenological fits to inclusive QE electron scattering data on different nuclei~\cite{Bodek19}, and also by fits to proton-nucleus scattering data~\cite{Cooper09}, in which the energy dependence is found to be similar for light and heavy systems. 
Thus, though in this work we present carbon results only, this method could be applied to other nuclei independently of its mass and isospin~\cite{Gonzalez-Jimenez19b}.

In what follows we refer to this approach as ED-RMF model.

\section{Results and Discussion}\label{Results}

In Sect.~\ref{re:e,e'}, we study Pauli blocking and distortion effects in $(e,e')$ cross sections. Since the energy and momentum transfer are known, contrary to the neutrino scattering case, it is easier to identify and separate the contributions from the different channels to the cross section. This simplifies the analysis and serves to benchmark the models.
We then study inclusive neutrino induced cross sections for fixed and flux-averaged neutrino energies in Sect.~\ref{re:nuenumu} and \ref{re:CCQE-MB}, respectively.
Finally, SPP cross sections are investigated in Sect.~\ref{re:SPP}. 
 
In addition to the QE and SPP contributions evaluated in this work, two-body current mechanisms, such as meson-exchange currents (MEC) and short-range correlations (SRC), also affect the 1p-1h and the 2p-2h responses. 
These have been previously evaluated in the framework of a nonrelativistic mean-field model, the effect of SRC was computed in Ref.~\cite{VanCuyck16} and the MEC in Refs.~\cite{Amaro92,Amaro94,VanCuyck17}. On the one hand, the influence of the two-body currents in the 1p-1h response was found to be very small. On the other hand, it is well known that the 2p-2h response plays a key role in the ‘dip’ region between the QE and delta peaks. Therefore, to compare with inclusive data we have included the 2p-2h MEC from Refs.~\cite{Megias15, DePace03} as a separate contribution. This consists in a fully relativistic microscopic calculation of one-pion exchanged two-body currents within a relativistic Fermi gas (RFG) model. 
In Ref.~\cite{Megias15}, it is argued that the general behavior of the MEC response is rather insensitive to finite-size nuclear effects, being dominated by the nucleon and pion electromagnetic form factors and the two particle-phase space. In spite of that, it would be very interesting to study two-body current mechanisms in a consistent way from the RMF theoretical stand-point, and evaluate if they differ much from those computed within nonrelativistic mean-field models and the RFG.

\subsection{Inclusive electron scattering}\label{re:e,e'}

\begin{figure*}[htbp]
  \centering
      \includegraphics[width=.23\textwidth,angle=270]{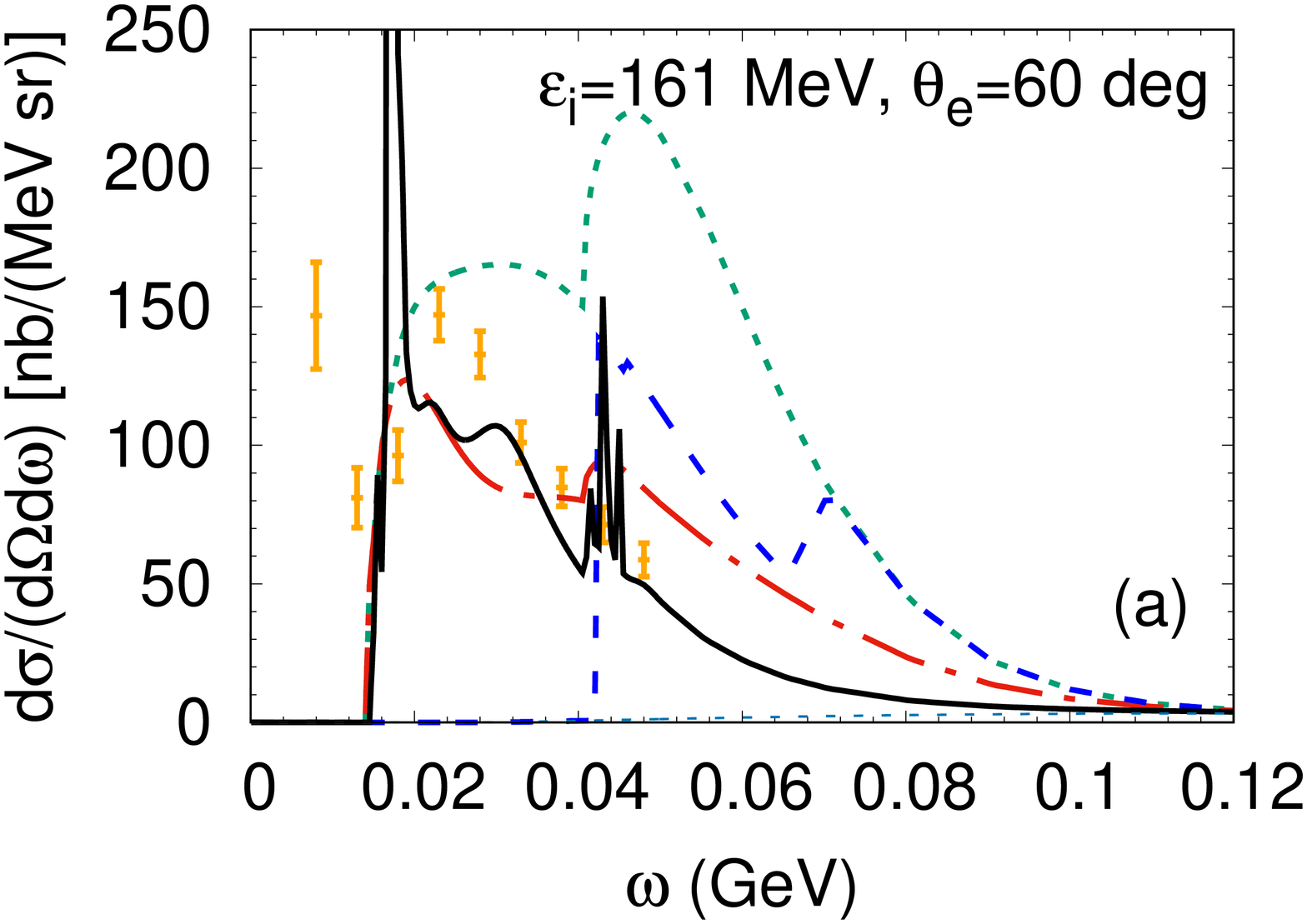}
      \includegraphics[width=.23\textwidth,angle=270]{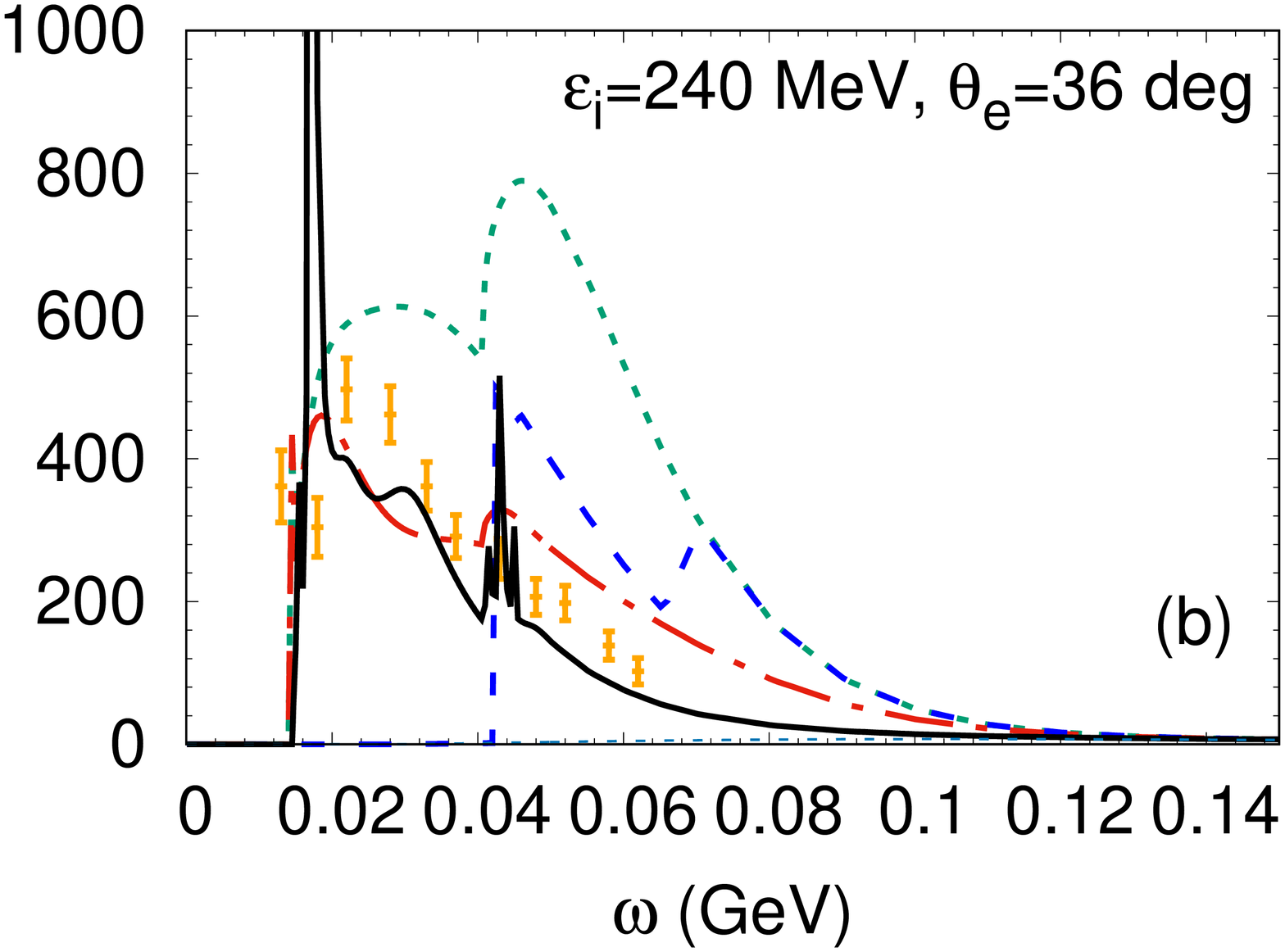}
      \includegraphics[width=.23\textwidth,angle=270]{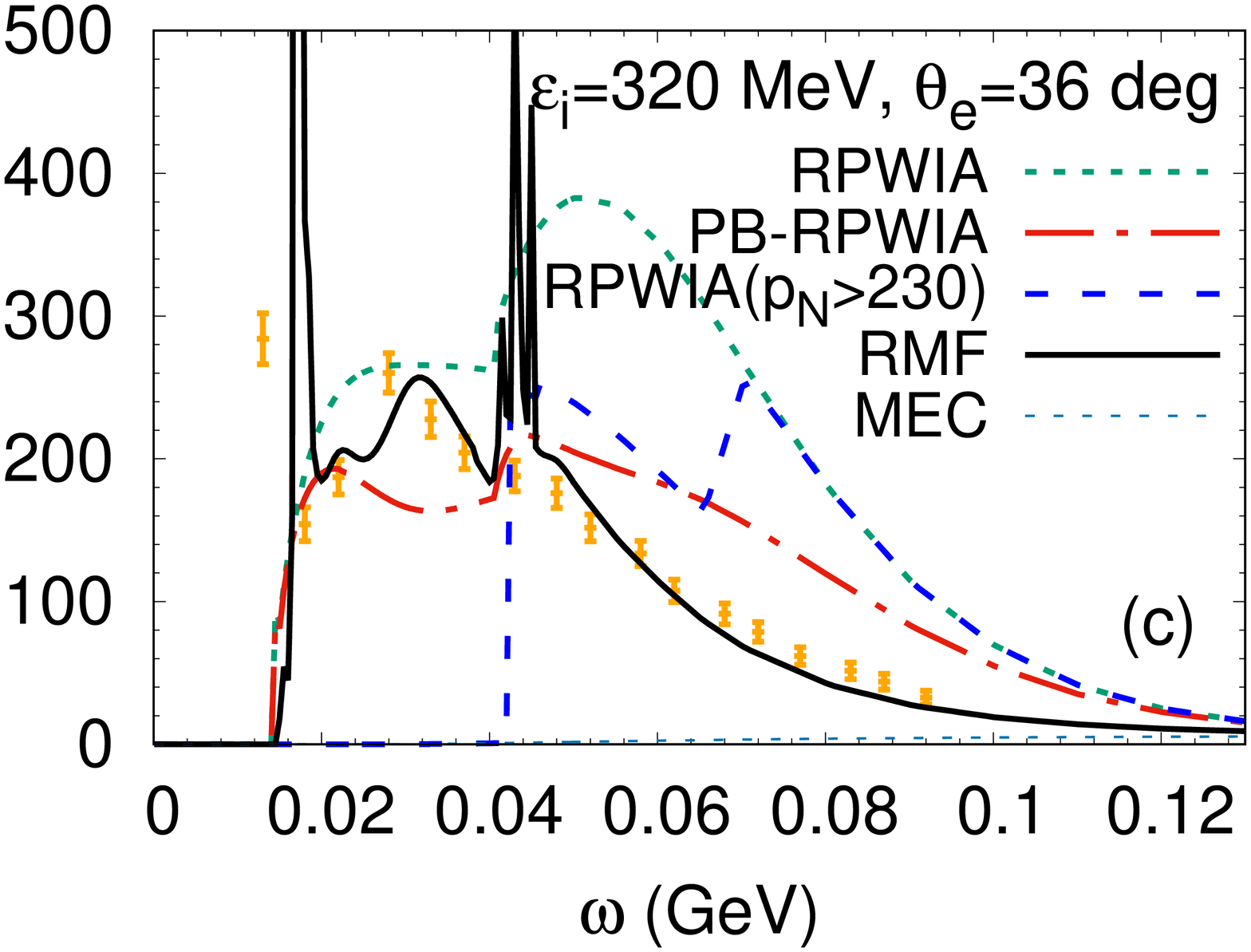}
      \caption{QE predictions for the process $^{12}C(e,e')$ with the RPWIA, PB-RPWIA, RPWIA($p_N>230$), and RMF models. The MEC contribution~\cite{Megias15} is shown separately. Although it is negligible at these kinematics, it has been added to the QE response. $\varepsilon_i$ is the incident electron energy and $\theta_e$ the scattering angle. Data taken from~\cite{ee-data}.}
  \label{fig:ee1} 
\end{figure*}

\begin{figure*}[htbp]
  \centering
      \includegraphics[width=.3\textwidth,angle=270]{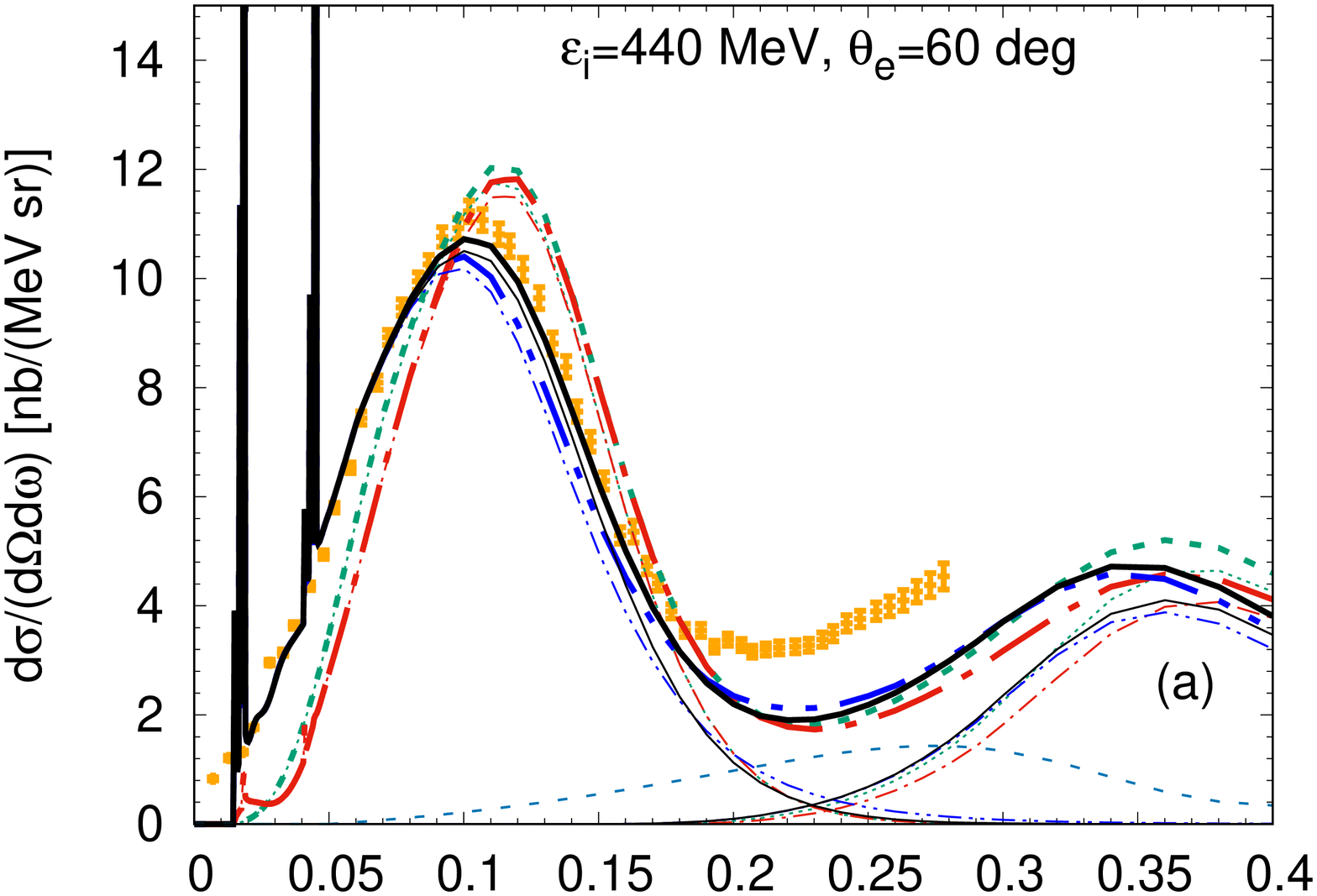}
      \includegraphics[width=.3\textwidth,angle=270]{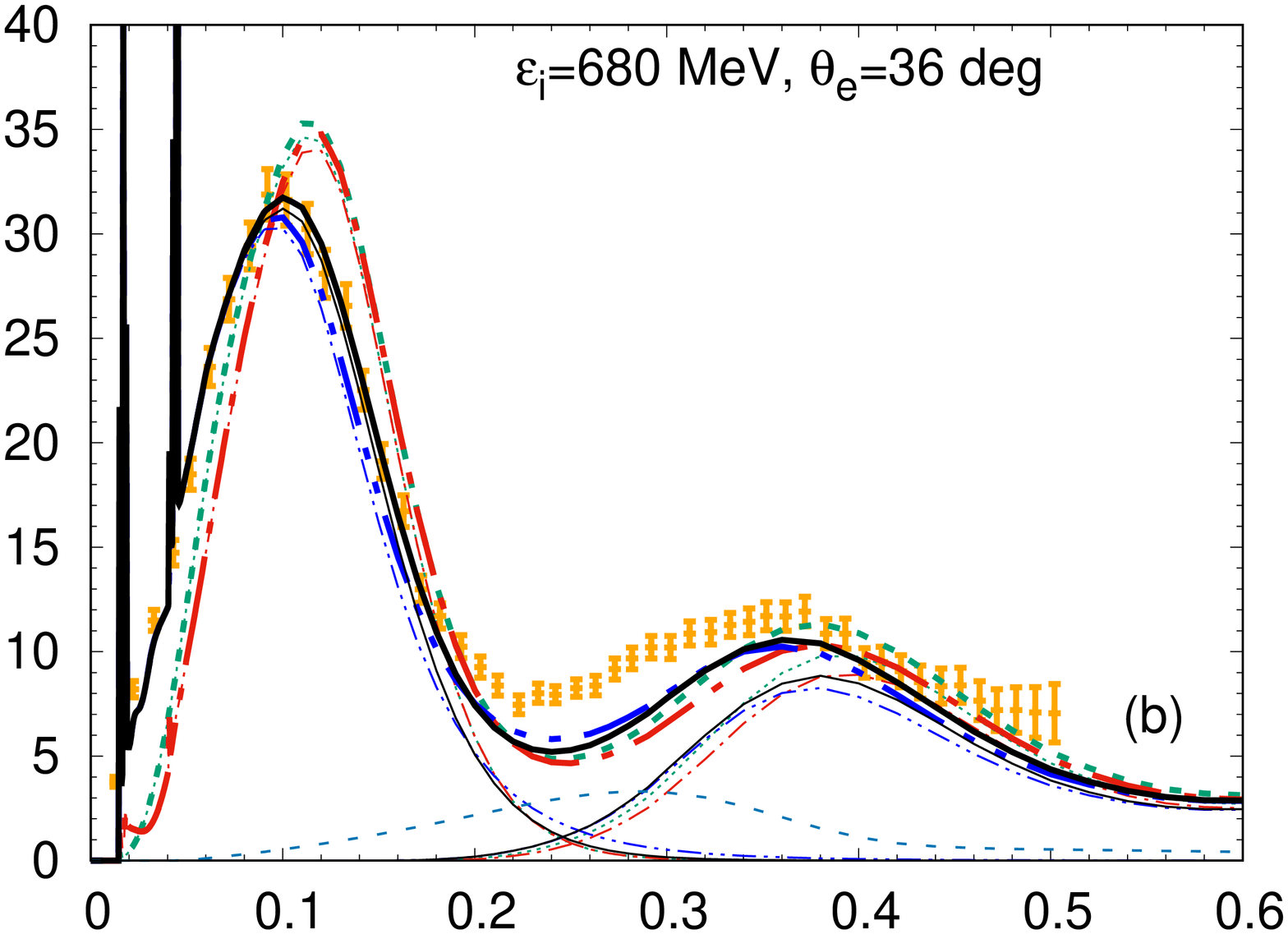}\\
      \includegraphics[width=.3\textwidth,angle=270]{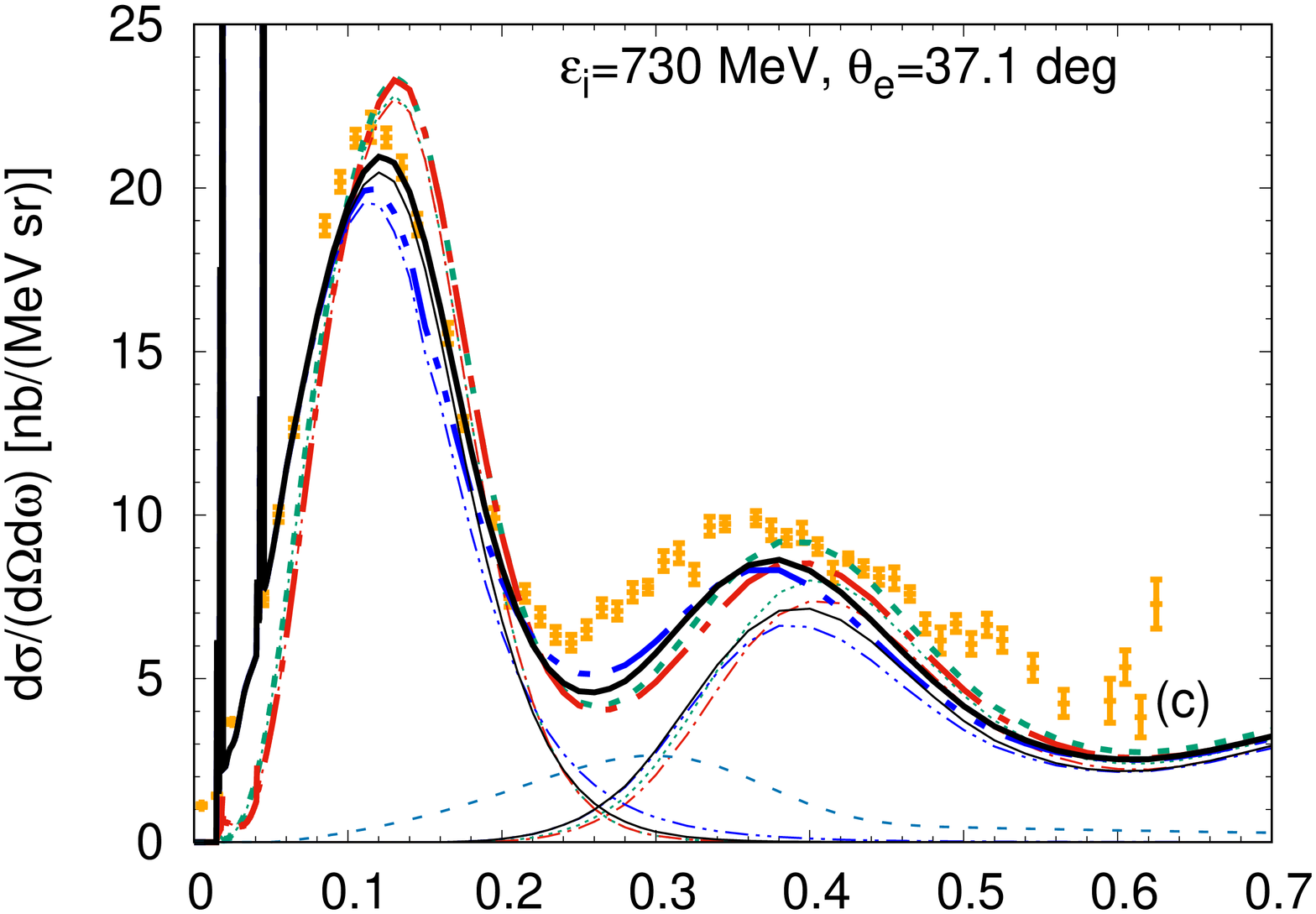}
      \includegraphics[width=.3\textwidth,angle=270]{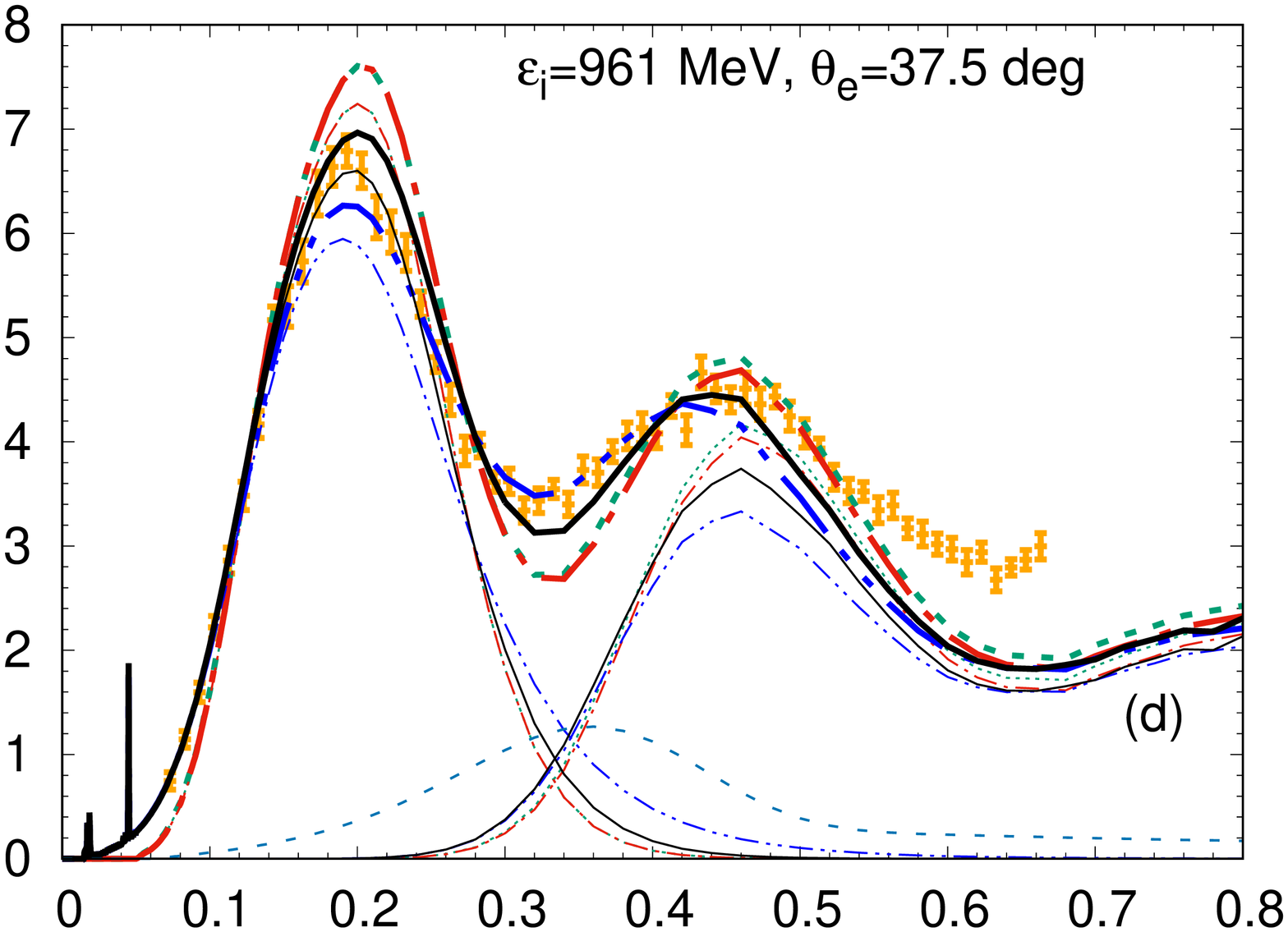}\\
      \includegraphics[width=.3\textwidth,angle=270]{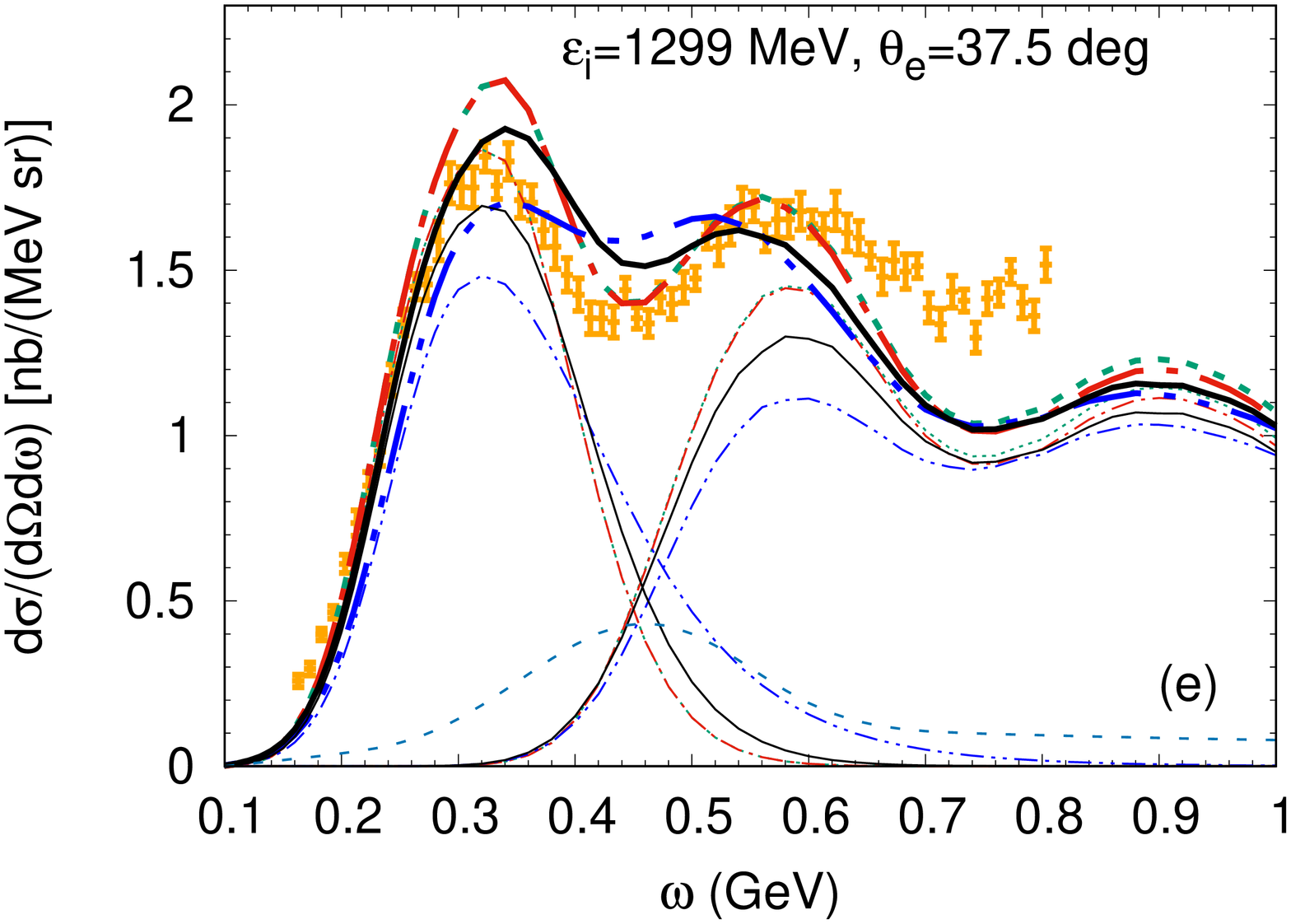}
      \includegraphics[width=.3\textwidth,angle=270]{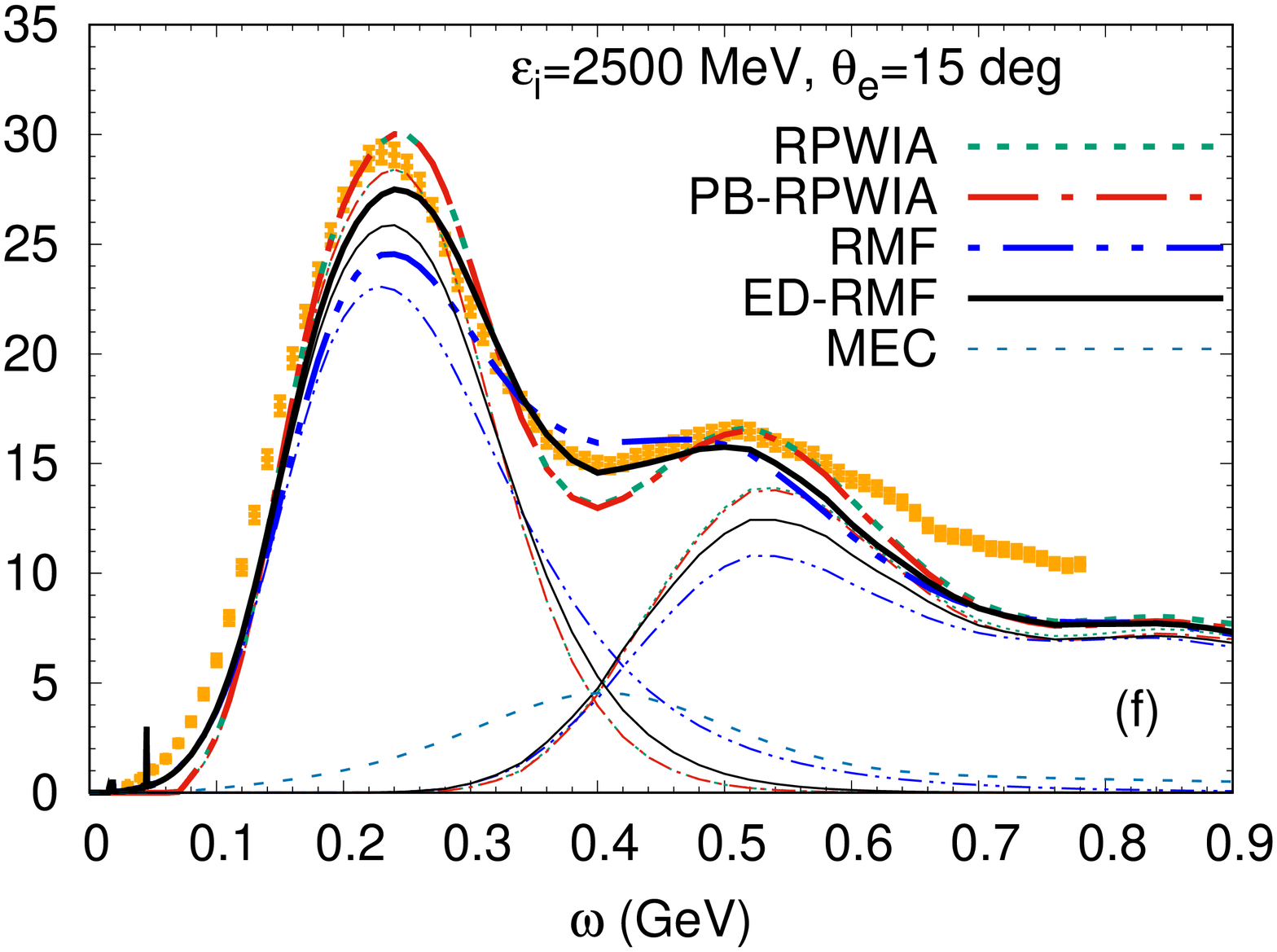}
  \caption{ QE and SPP  for the process $^{12}C(e,e')$ with the RPWIA, PB-RPWIA, RMF and ED-RMF models. MEC was taken from Ref.~\cite{Megias15}. The QE+MEC+SPP cross sections are represented by the thicker lines, while the QE and SPP cross sections correspond to the thinner lines. Data from~\cite{ee-data}.}
  \label{fig:ee2}  
\end{figure*}

In Fig.~\ref{fig:ee1} and \ref{fig:ee2}, we compare the results from the different models with $^{12}C(e,e')$ data. 

For the kinematics of Fig.~\ref{fig:ee1}, well below the pion production threshold, one-nucleon knockout (QE scattering) and collective nuclear effects~\cite{Pandey15} dominate the nuclear response. The latter are not explicitly included in our calculations so we expect some underestimation of the data. The too narrow resonances that appear in the RMF results would certainly smooth out in a beyond mean-field approach including the effect of the residual interactions.
The effect of Pauli blocking is analyzed by comparing RPWIA (dashed-green lines) to PB-RPWIA (dashed-dot-red lines). One observes that RPWIA overshoots the data. Pauli blocking removes the spurious strength coming from the non-orthogonality of the hadronic states, which considerably improves the comparison with data. 

One may be tempted to take a short cut and introduce `Pauli blocking' using the same procedure as in Fermi-gas based models, that is, through a simple cutoff that eliminates the contributions corresponding to knockout nucleons with momenta below a given Fermi momentum $p_F$ (we use $p_F=230$ MeV for carbon). The results of this approach are shown by the dashed-blue lines. This approach seems to provide the right total strength (area under the curves) though the position of the cross section is clearly off.
Notice that the binding energy is already included, in the same way, in all models shown here, so there is no freedom left to shift the position of the distributions. 

The effect of the distortion of the outgoing nucleon can be inferred by comparing PB-RPWIA and RMF results. Still in Fig.~\ref{fig:ee1}, one observes a further reduction and redistribution of the strength that improves the agreement with data. For the kinematics of Fig.~\ref{fig:ee1}, the ED-RMF approach provides the same results as the RMF model so we do not show the results here. 
One could say that, for low momentum of the outgoing nucleus, the mere restoration of orthogonalization is sufficient to bring the RPWIA results in good agreement with the data. However, when one reaches the 'tail' of the response at higher energy transfers, distortion of the final nucleon, that is, the fact that the momentum of the knockout nucleon inside the nucleus is not coincident with the momentum of the detected nucleon (asymptotic value) but it is smeared out by FSI, is fundamental to obtain agreement with the data. Thus the RMF is clearly favoured by the data.
Similar results are obtained with the mean-field approach of Ref.~\cite{Pandey15}. 

The influence of the distortion is best seen in Fig.~\ref{fig:ee2}. For these kinematics, the QE and delta peaks are clearly defined. Also, the contribution of two-nucleon knockout, as estimated in~\cite{Megias15}, is important. We discuss first the QE contribution. One observes that Pauli blocking has relatively little impact while the distortion strongly shifts the distributions producing an excellent agreement with the data, especially in the $\omega$ region below the QE peak.
Regarding the SPP cross sections, in Figs.~\ref{fig:ee2} (a), (b) and (c) the effect of Pauli blocking shows as an overall reduction of the order of $10\%$ in the peaks. The distortion produces a small shift of the distributions towards lower $\omega$. In Fig.~\ref{fig:ee2} (d), (e) and (f) Pauli blocking produces a small effect while the distortion considerably reduces the cross section.

Notice that, since we are comparing with inclusive samples, it is expected (and desirable) to underestimate the data, especially in the region above the pion production threshold, where other processes beyond SPP, such as two-pion production~\cite{Nakamura15}, start to contribute. 
Also, it is convenient to clarify that we do not include medium-modification of the delta width (MM). 
We have verified that, for the kinematics of Fig.~\ref{fig:ee2}, the model for the MM of Refs.~\cite{Oset87,Salcedo88,Nieves93} leads to a reduction of the cross section in the delta region of approximately $15-20\%$. This is due to the new decay channels that are open: $\Delta N\rightarrow \pi NN$, $\Delta N\rightarrow NN$ and $\Delta NN\rightarrow NNN$. In order to reproduce the inclusive signal, those processes should be modeled and their contributions added to the cross section, what might compensate the reduction.

\begin{figure}[htbp]
  \centering
      \includegraphics[width=3.2cm,height=3.8cm,angle=270]{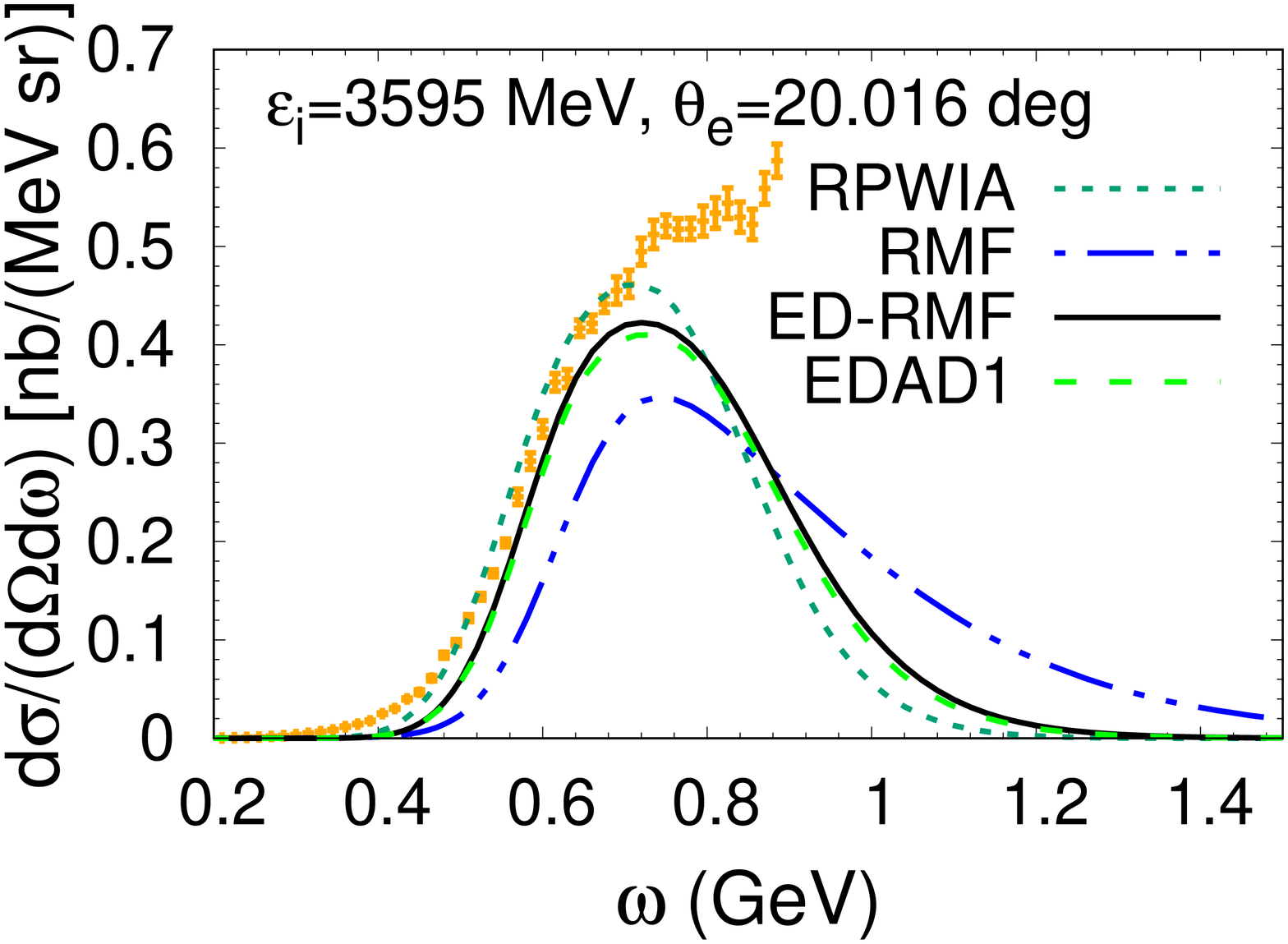}
      \includegraphics[width=3.2cm,height=3.6cm,angle=270]{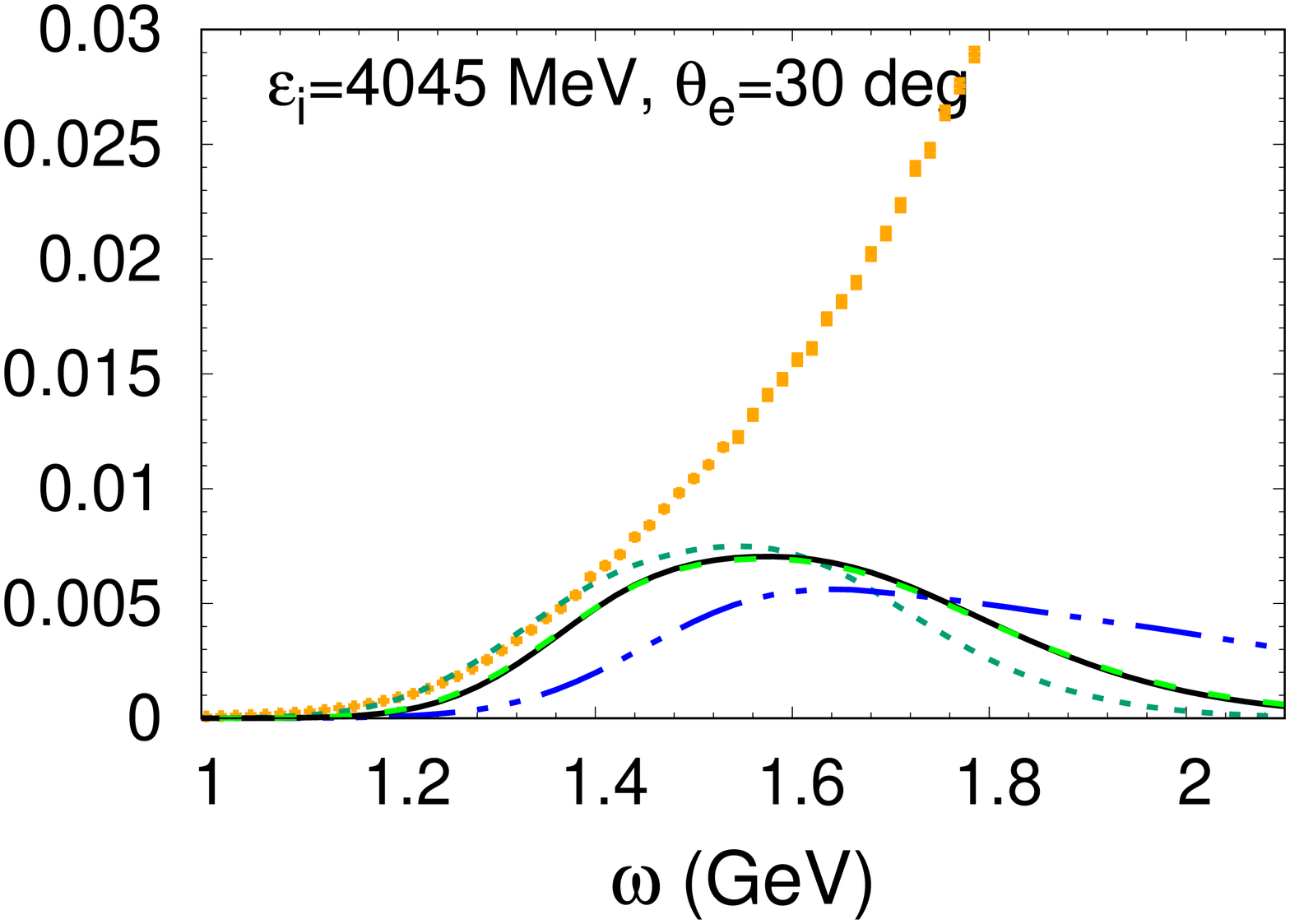}
  \caption{ QE contribution computed with RPWIA, RMF and ED-RMF models. Data taken from~\cite{ee-data}.}
  \label{fig:ee3}  
\end{figure}

In Fig.~\ref{fig:ee2} we have also included the results from the ED-RMF model. We observe the expected and desired effect, i.e., the ED-RMF results lay in between the RPWIA and RMF ones, improving the agreement with data. To further illustrate this, in Fig.~\ref{fig:ee3} we show the QE results from the RPWIA, RMF and ED-RMF models for two extreme high-energy kinematics. 

\subsection{Inclusive neutrino scattering}\label{re:nuenumu}

Here we present neutrino induced QE cross sections for fixed incoming energies. A situation which is currently unattainable in experimental situations but important to study for two main reasons. First, it allows for a clear separation of different reaction processes, and second, the exact shape of the double differential cross section for fixed incoming energies is a crucial ingredient in the analysis of neutrino oscillations as it links the experimentally obtainable information to the neutrino energy~\cite{Nikolakopoulos18b}.

\begin{figure}
\includegraphics[width=0.23\textwidth,angle=270]{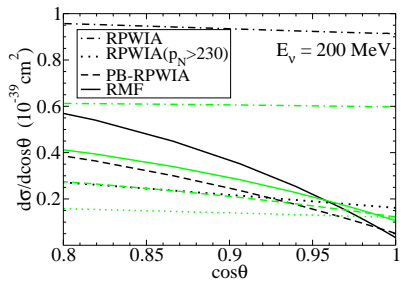}
\caption{QE single-differential cross section as a function of the scattering angle. Dark-black and soft-green lines correspond to electron and muon-neutrino induced reactions, respectively. Within the RMF and PB-RPWIA approaches the electron-neutrino cross section become smaller than the muon-neutrino one for very forward scattering angles.}
\label{fig:numu-nue}
\end{figure}

In Refs.~\cite{Martini16,Nikolakopoulos19}, it was shown that when including the distortion of the outgoing nucleon wave functions, for forward scattering of the charged lepton, the cross section for quasielastic nucleon knockout is larger for muon neutrinos than for electron neutrinos. We further analyze this here. In Fig.~\ref{fig:numu-nue} we show the electron- and muon-neutrino single-differential cross section as a function of scattering angle. One sees that within the RPWIA and the `RPWIA with the cutoff' the electron-neutrino cross section is always larger than the muon-neutrino one. Contrary, the RMF and PB-RPWIA predict a larger muon-neutrino cross section for very forward scattering angles. This suggests that a proper quantum mechanical treatment of the Pauli blocking is the key to reproduce this effect~\cite{Nikolakopoulos19}.

\begin{figure}
\includegraphics[width=0.35\textwidth]{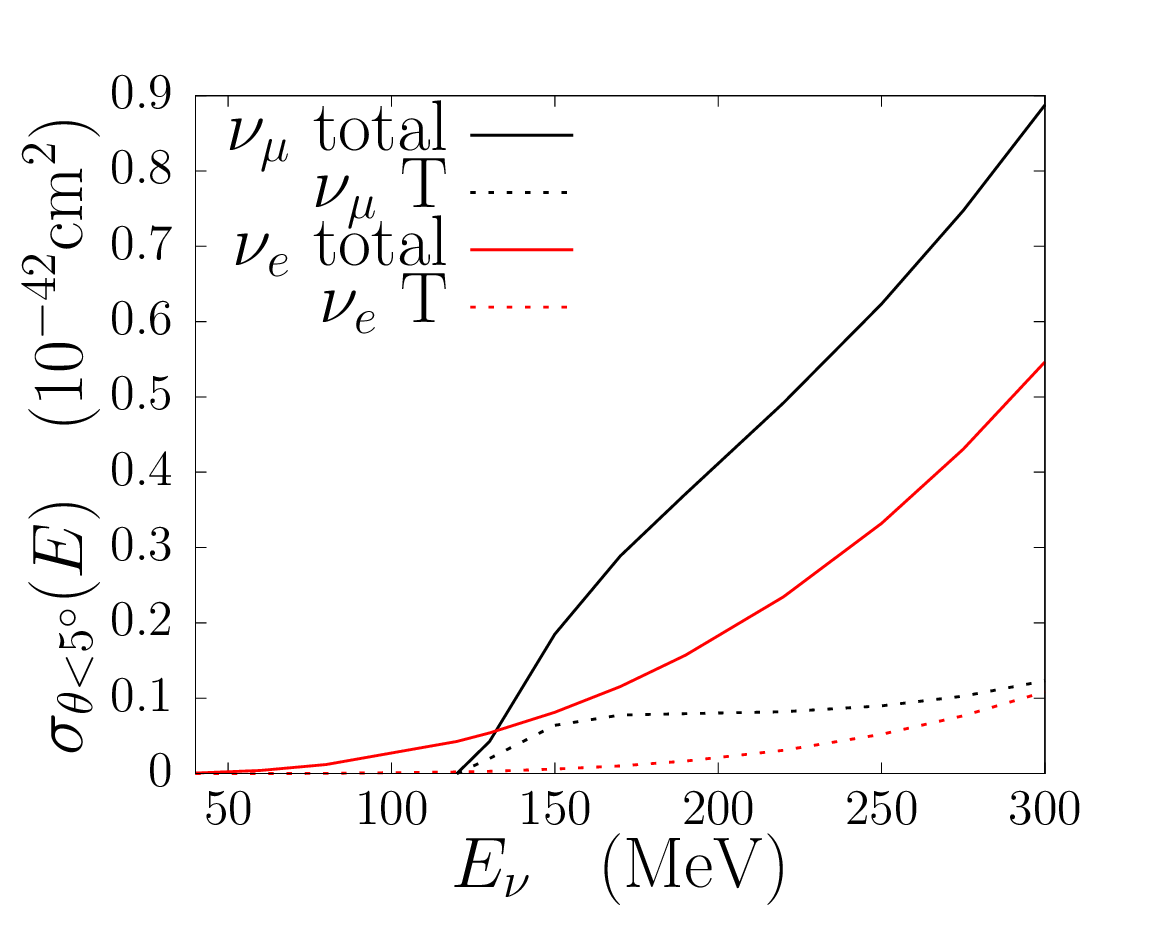}
\caption{QE total cross section per neutron restricted to scattering angles smaller than 5 degrees. For low energies around the threshold the muon-induced cross section receives a non negligible contribution from the transverse response. Calculations with the RMF model.}
\label{fig:totCSLandT}
\end{figure}

In Fig.~\ref{fig:totCSLandT} we show the QE total cross section for interactions restricted to very forward scattering angles, where the total transverse response is shown separately to highlight a peculiarity of the interaction which is important around the muon threshold and at very forward scattering angles.
In the massless limit for forward lepton scattering the transverse lepton kinematical factors are zero. For the heavier muon, however, the transverse responses are non negligible when the muon mass is large compared to its momentum.

\begin{figure}
\includegraphics[width=0.5\textwidth]{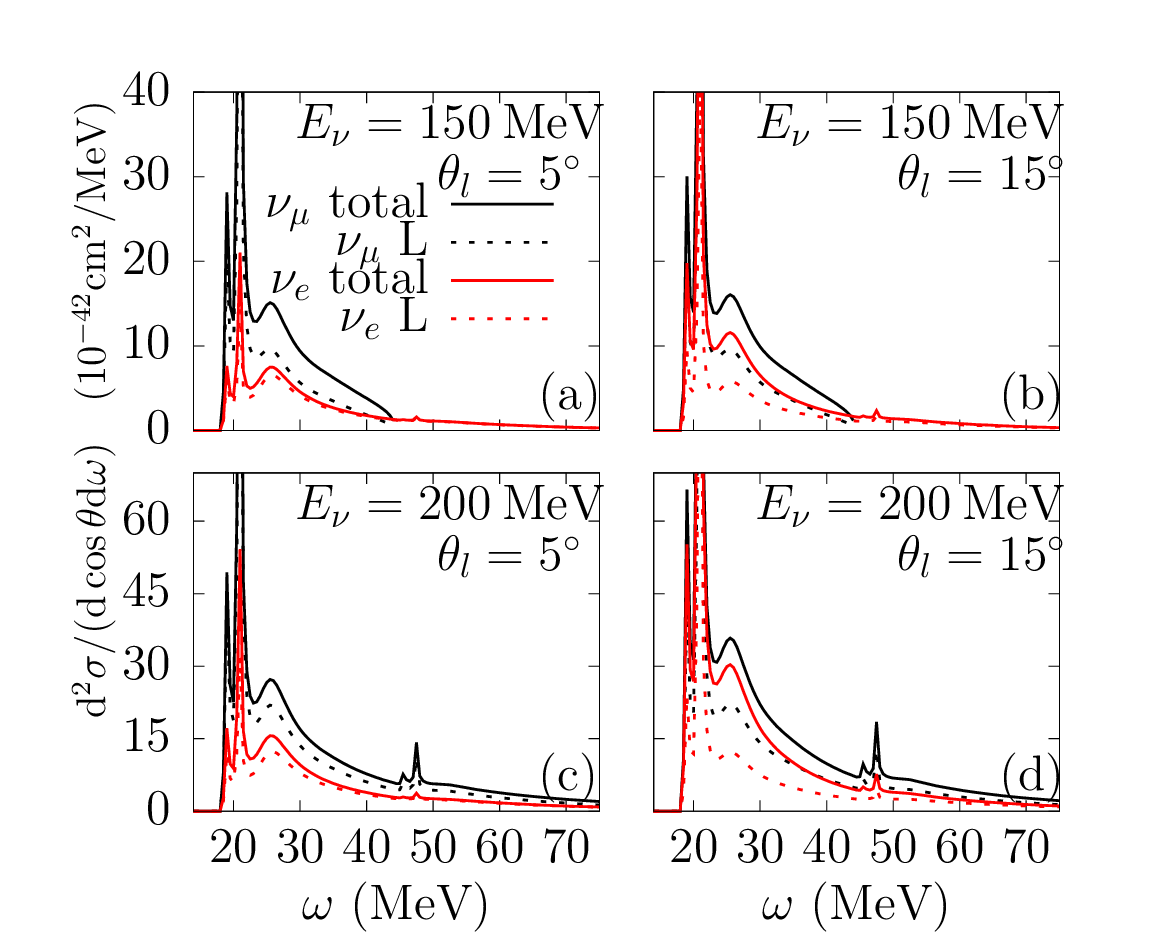}
\caption{Double differential cross section for electron and muon neutrino quasi-elastic scattering on carbon. The longitudinal contribution to the cross section is shown separately. Calculations performed with the RMF model. }
\label{fig:LandT}
\end{figure}

Fig.~\ref{fig:LandT} depicts the double differential cross section for forward lepton scattering and energies near the muon production threshold. The longitudinal contribution is shown separately, so that the difference between the total (solid lines) and the longitudinal contributions (dashed lines) is the total transverse (T and T') cross section.
In Fig.~\ref{fig:LandT}(a), we see indeed that the electron neutrino induced cross section is almost completely longitudinal while the transverse response gives an important contribution to the $\nu_\mu$ cross section.
For slightly larger incoming energies but still very forward scattering, Fig.~\ref{fig:LandT}(c), we see that both the muon and electron cross sections are primarily longitudinal.
For larger scattering angles, Figs.~\ref{fig:LandT}(b) and (d), the transverse cross section becomes more important. 
A similar analysis, with similar outcomes, was previously presented in Ref.~\cite{Martini16} using a different model, the continuum random phase approximation (CRPA).

\begin{figure}[htbp]
  \centering
      \includegraphics[width=.25\textwidth,angle=270]{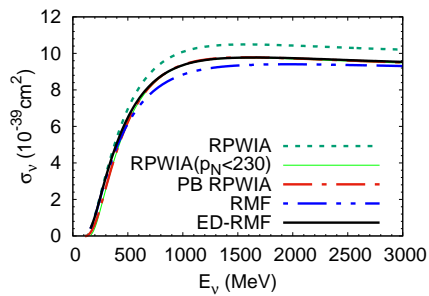}
  \caption{ Total cross section per neutron for CCQE $\nu_\mu-^{12}C$. }
  \label{fig:total-CCQE}  
\end{figure}

Finally, in Fig.~\ref{fig:total-CCQE} we show the total CCQE cross sections for the different models. It is interesting that from approximately $E_\nu>700$ MeV, the PB-RPWIA, RPWIA with the cutoff and the ED-RMF overlap, laying in between the RPWIA and RMF results, which could be considered as an upper and lower bound for the inclusive cross section.
  
\subsection{MiniBooNE CCQE muon-neutrino scattering}\label{re:CCQE-MB}

To analyze the nuclear effects on flux-averaged neutrino-nucleus distributions, we study here the CCQE cross sections for the muon-neutrino MiniBooNE flux~\cite{MiniBooNECC10}.
 
\begin{figure}[htbp]
  \centering
      \includegraphics[width=.33\textwidth,angle=270]{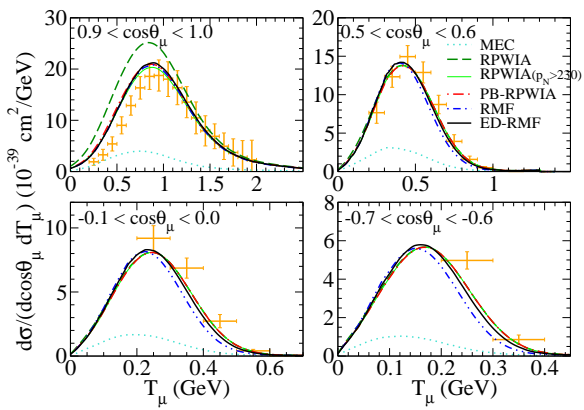}      
  \caption{Double differential CCQE cross section folded with MiniBooNE flux. The MEC contribution from~\cite{Megias16b} is shown separately and has been added to all QE results. Data from~\cite{MiniBooNECC10}.}
  \label{fig:ds2-dTdcos}  
\end{figure}

\begin{figure}[htbp]
  \centering
      \includegraphics[width=.25\textwidth,angle=270]{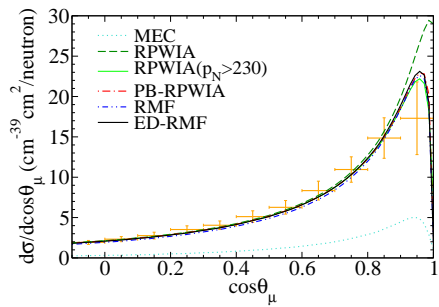}
  \caption{ Single differential CCQE cross sections folded with the MiniBooNE flux. MEC, taken from~\cite{Megias16b}, is shown separately and has been added to all QE results. Data from~\cite{MiniBooNECC10}.}
  \label{fig:ds-dcos}  
\end{figure}

The CCQE differential cross section in bins of $\cos\theta_\mu$ as a function of the muon kinetic energy ($T_\mu$) is represented in Fig.~\ref{fig:ds2-dTdcos}. The effect of Pauli blocking is visible only for the most forward bin of Fig.~\ref{fig:ds2-dTdcos} (a), which was expected since forward scattering cross sections are dominated by events where the neutrino transfers small energy and momentum. The effect of the distortion (RMF vs PB-RPWIA) is visible for the less forward bins [Figs.~\ref{fig:ds2-dTdcos} (b), (c) and (d)] and consists in an important reduction of the cross section on the right-hand side of the peak. This effect, however, disappears when the ED-RMF approach is employed. 
This is consistent with the results for the single differential cross section shown in Fig.~\ref{fig:ds-dcos}. In this case, due to the integration over the muon energy, the PB-RPWIA result almost matches the ED-RMF one. The main difference with respect to RPWIA is an important reduction of the cross sections at forward angles. 

\subsection{CC neutrino-induced SPP}\label{re:SPP}

In this section we present results for CC neutrino-induced one-$\pi^+$ production on carbon, and show how the description of the knockout nucleon wave function affects the cross section.
All the results in this section include medium-modification of the delta decay width~\cite{Oset87,Salcedo88,Nieves93}.

\begin{figure}
\includegraphics[width=0.27\textwidth,angle=270]{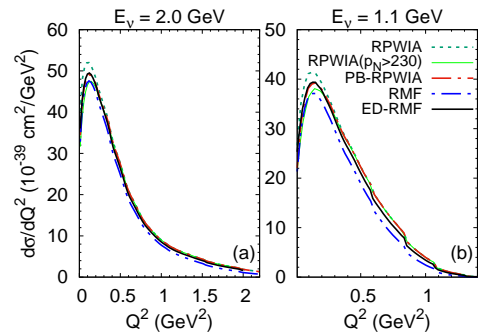}
\caption{Single charged pion production on carbon. Cross section in terms of $Q^2$ for two neutrino energies.}
\label{fig:SPPQQ}
\end{figure}

In Fig.~\ref{fig:SPPQQ} we show the single differential cross section as a function of $Q^2$, for two different incoming energies.
The $Q^2$ distribution is a topic of current debate in the neutrino community. In recent experiments it was found that a suppression of the resonance production cross section at low $Q^2$ is necessary to obtain agreement with the data~\cite{MINERvA-QElike-19}.
The reason for this suppression is however not understood.
Looking at these results we see that the inclusion of Pauli blocking leads to a reduction in precisely this low $Q^2$ region (compare RPWIA to PB-RPWIA lines) while the effect of the distortion is almost averaged out as inferred from the comparison ED-RMF to PB-RPWIA. 
Therefore, since MC event generators do have Pauli blocking implemented in a way which is somehow similar to the `RPWIA($p_N>230$)' approach, with the results we have at hand, we cannot ascribe the experimentally suggested quenching of the resonance production yield to nucleon final state effects.

\begin{figure}
\includegraphics[width=0.25\textwidth,angle=270]{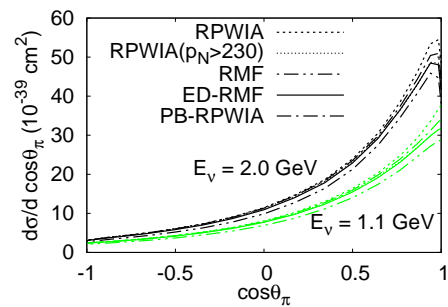}
\caption{Single charged pion production cross section on carbon as a function of $\cos\theta_\pi$ for $E_\nu = 2$ GeV (dark-black lines) and $E_\nu = 1.1$ GeV (soft-green lines).}
\label{fig:ct_pi}
\end{figure}

The angular distribution of the pion is shown in Fig.~\ref{fig:ct_pi}. The PB-RPWIA leads to a reduction of the cross section for forward scattering angles. This can be understood from the fact that forward scattered pions generally have higher energy than those scattered backwards, this means that the outgoing nucleon in this kinematic region has low momentum and therefore has a non negligible overlap with the initial state. 
This reduction is of course also present in the RMF calculation, but the RMF result shows an additional reduction of scattering strength over the whole angular space.
As described in the previous sections, the predictions from PB-RPWIA and ED-RMF approaches lay in between RPWIA and RMF ones, which could be considered as an upper and lower bound, respectively.

\section{Conclusions}\label{Conclusions}

In scattering reactions, the interaction of the particles in the final-state with the residual nucleus affects the cross sections regardless of whether these particles are detected or not. This quantum mechanical effect does not occur in classical or semi-classical approaches which instead perform a trivial factorization of the reaction vertex and the propagation of the outgoing particles. 
This is of relevance for the neutrino community since, often, the signal in the detector consists of, e.g., a lepton (muon or electron) detected and nothing else. A correct prediction of the lepton distributions, therefore, requires proper treatment of the elementary vertex and the hadronic final-state interactions.

In this work, we have studied the effects on the quasielastic (QE) and single-pion production (SPP) cross sections that arise from the differences in the treatment of the knockout nucleon wave function.

The influence of Pauli blocking (PB) has been studied by comparing the results from three approaches: RPWIA, PB-RPWIA and RPWIA with a cutoff. Within RPWIA, the outgoing nucleon is a relativistic plain wave, thus FSI (including the PB) are ignored. This results in an clear overestimation of data in the low-$\omega$ region (Fig.~\ref{fig:ee1}) and a wrong prediction of the position of the QE peak (Fig.~\ref{fig:ee2}). After the flux folding for neutrino-induced reactions, the extra strength appears at forward scattering angles, see Figs.~\ref{fig:ds2-dTdcos}(a) and \ref{fig:ds-dcos}. 
In analogy to the procedure followed in Fermi-gas based models, one can introduce a cutoff for the knockout nucleons with momentum below a given Fermi momentum $p_F$. This is the easiest way to somehow account for the PB. This approach is correct and consistent for infinite nuclear matter (global Fermi gas model), but not in more realistic frameworks like shell models, where the nucleons are labeled by energy and angular momentum quantum numbers rather than with their momenta. In Fig.~\ref{fig:ee1} it is shown that this cutoff leads to approximately the right total strength but the position of the distributions is clearly off.
Within the PB-RPWIA approach, the PB is incorporated in a way which is consistent with our relativistic quantum mechanical framework. Basically, we make the initial and final states orthogonal by subtracting their overlap, hence, spurious contributions are eliminated. Contrary to `RPWIA with a cutoff', this approach allows nucleons with low momentum to leave the nucleus, what results in a more realistic shape of the low energy cross sections. 
Additionally, we have shown that a correct treatment of the PB seems to be the key to predict the right ratio of muon- to electron-neutrino cross sections at very forward scattering angles (see Fig.~\ref{fig:numu-nue} and Ref.~\cite{Nikolakopoulos19} for a more detailed discussion).

The distortion of the outgoing-nucleon wave functions due to the presence of the residual nucleus have been studied by comparing the former approaches with the RMF and the ED-RMF models.
Within the RMF the initial and final nucleons are eigenstates of the same hamiltonian, so PB and distortion are naturally incorporated in a consistent way. First, one observes an important shift of the distributions. Second, there is a redistribution of the strength towards the tails. Both effects tend to improve the agreement with the inclusive electron scattering data (Fig.~\ref{fig:ee1} and \ref{fig:ee2}), as long as the knockout nucleon has a momentum approximately below $500$ MeV. 
In the ED-RMF approach, introduced for the first time in this work, we have replaced the RMF potentials felt by the outgoing nucleon by energy-dependent potentials. This is done by scaling down the scalar and vector RMF potentials as the momentum of the nucleon increases. 
This introduces in a phenomenological way the right behaviour of the model for high energy transfer, while keeping the RMF limit for nucleons with small momenta, naturally incorporating orthogonality and FSI.

It is interesting that after folding over the neutrino energy, in particular for the MiniBooNE flux and the CCQE samples studied here (Figs.~\ref{fig:ds2-dTdcos} and \ref{fig:ds-dcos}), all approaches provide very similar results except the RPWIA, which is systematically larger. The effect of the nucleon distortion seems to average out in flux-folded distributions, meaning that the reduction of the cross section implied by Pauli blocking is quantitatively the most relevant effect, but it can be reproduced with a simple cutoff approach.
In spite of that, there are important differences in the double differential cross section, as shown in Sect.~\ref{re:e,e'}. These differences in the cross section can affect the oscillation analyses as they are the link between the data in terms of the final-lepton variables and the reconstructed neutrino energy distributions~\cite{Nikolakopoulos18b}.

In Figs.~\ref{fig:SPPQQ} and \ref{fig:ct_pi} we have shown the effect of PB and nucleon distortion on the neutrino-induced SPP cross sections. In particular, we have analyzed the $Q^2$ and pion scattering angle distributions. The effects are similar to those observed and discussed above, i.e. a reduction and shift of the distributions. 
This corroborates that the treatment of final state particles is important regardless of whether these are part of the signal or not.
We want to stress that, to our knowledge, this is the first time that the distortion of the knocked out nucleon is implemented for neutrino-induced SPP on the nucleus within a fully relativistic and quantum mechanical mean field framework.

The effects linked to the distortion of the pion wave function will be studied in future work. Two-body current mechanims, affecting the 1p-1h and 2p-2h responses, should also be computed in a consistent RMF framework in the light of a more complete model.

\begin{acknowledgments}
We thank K. Niewczas for useful discussions in many stages of this work. 
R.G.J. sincerely thanks T.W. Donnelly for the conversations about the Pauli blocking held during the EINN-2017 conference (Cyprus), those inspiring ideas were seminal for part of this work.
This work was partially supported by the Interuniversity Attraction Poles Programme initiated by the Belgian Science Policy Office (BriX network P7/12) and the Research Foundation Flanders (FWO-Flanders) and Special Research Fund, Ghent University; and by Spanish Government (FPA2015-65035-P) and Comunidad de Madrid (B2017/BMD-3888 PRONTO-CM) and European Regional Funds.
R.G.J. acknowledges support by Comunidad de Madrid and UCM under the contract No. 2017-T2/TIC-5252.
The computations of this work were performed in EOLO, the HPC of Climate Change of the International Campus of Excellence of Moncloa, funded by MECD and MICINN as a contribution to CEI Moncloa; the high capacity cluster for physics, funded by UCM and FEDER funds, CEI Moncloa; and the Stevin Supercomputer Infrastructure provided by Ghent University, the Hercules Foundation and the Flemish Government.
\end{acknowledgments}

\appendix

\section{Analytic expressions of the projection coefficients} \label{proj-coeff}

The projection coefficients $C^{m_j,s_N}_{\kappa}(\np_N)$ are defined as
\fl{
  C^{m_j,s_N}_{\kappa}(\np_N) &= \int_V d\nr\ [\psi_{pw}^{s_N} (\nr,\np_N)]^\dagger\, \psi_\kappa^{m_j}(\nr)  \\
  &= \sqrt{\frac{(2\pi)^{3}M}{V E_N}} \uuN^\dagger \psi_\kappa^{m_j}(\np_N)\,,\label{C}\non
}
where $\psi_\kappa^{m_j}(\np_N)$ is the Fourier transformed wave function of $\psi_\kappa^{m_j}(\nr)$. 
We can get analytic expressions for these coefficients $C^{m_j,s_N}_{\kappa}(\np_N)$. 

After some algebra one obtains:
\fl{
  C^{m_j,s_N}_{\kappa}(\np_N) = \frac{1}{\sqrt{V}} \eta_\kappa(p_N) \left[\chi_{s_N}^\dagger\ \varphi_\kappa^{m_j}(\Omega_{\np_N})\right]\,,
}
with 
\ba
  \eta_\kappa(p_N) &=& (2\pi)^{3/2}\sqrt{\frac{M}{E_N}}\, (-i)^\ell \label{eta}\\
  &\times&\left(g_\kappa(p_N)+S_\kappa f_\kappa(p_N)\frac{p_N}{E_N+M}\right)\,,\non
\ea
and
\fl{
\chi_+^\dagger\ \varphi_\kappa^{m_j}(\Omega_{\np_N}) &= \langle \ell(m_j-\half),\half\half |jm_j\rangle \non\\
 &\times Y_\ell^{(m_j-\half)}(\Omega_{\np_N})\,, \label{proj+}
 }
\fl{
\chi_-^\dagger\ \varphi_\kappa^{m_j}(\Omega_{\np_N}) &= \langle \ell(m_j+\half),\half\mhalf|jm_j\rangle\non\\
 &\times Y_\ell^{(m_j+\half)}(\Omega_{\np_N})\,. \label{proj-}
}
In Eq.~\ref{eta} $S_\kappa$ is the sign of $\kappa$, and $g_\kappa$ and $f_\kappa$ are the radial functions associated with the upper and lower components in the bound nucleon wave function in momentum space~\cite{Caballero98a}.

\section{Normalization of the Pauli blocked wave function}\label{normalization}

The Pauli blocked wave function $\Psi^{s_N}(\nr,\np_N)$, defined in Eq.~\ref{psi_block}, is correctly normalized to one. This is shown as follows: 
\fl{ 
 |\Psi^{s_N}(\np_N)|^2 = 1- \sum_{\kappa,m_j} |C^{m_j,s_N}_{\kappa}(\np_N)|^2 \equiv N\,,
}
with $N$ the norm. On the other hand, from Eq.~\ref{C} we have:
\fl{
  |C^{m_j,s_N}_{\kappa}(\np_N)|^2 = (2\pi)^{3}\frac{M}{VE} |\uuN^\dagger \psi_\kappa^{m_j}(\np_N)|^2\,.
}
Since the normalization volume $V$ can be taken arbitrarily large, and the quantity $(2\pi)^{3}\frac{M}{E}|\uuN^\dagger \psi_\kappa^{m_j}(\np_N)|^2$ is finite, in the limit $V\rightarrow\infty$, one gets 
$|C^{m_j,s_N}_{\kappa}(\np_N)|^2\rightarrow0$, so $N\rightarrow1$, as it should be.

\bibliographystyle{apsrev4-1}
\bibliography{bibliography}

\end{document}